\documentclass[twocolumn,a4paper]{article}
\usepackage[a4]{draft}

\usepackage{color}
\usepackage{cite}
\usepackage{amsmath,amssymb,amsfonts}
\usepackage{algorithmic}
\usepackage{graphicx}
\usepackage{fancybox}
\usepackage{multirow}
\usepackage{booktabs}

\begin{document}

\date{}

\title{\large\bf
Hardware-Efficient Accurate 4-bit Multiplier for Xilinx 7 Series FPGAs}

\author{
  \normalsize
  \begin{tabular}[t]{c@{\extracolsep{8em}}c}
    \large Misaki Kida & \large Shimpei Sato\\
    Graduate School of Science and Technology,  & Faculty of Engineering, \\
    Shinshu University & Shinshu University \\
    4-17-1 Wakasato, Nagano-city, Nagano, Japan & 4-17-1 Wakasato, Nagano-city, Nagano, Japan\\
    25w6030k@shinshu-u.ac.jp & satos@shinshu-u.ac.jp
  \end{tabular}
}

\maketitle

\thispagestyle{empty}

{\small\bf Abstract---
As IoT and edge inference proliferate, there is a growing need for lookup-table (LUT)-based multipliers that combine a low LUT count with a high maximum operating frequency. This paper proposes a hardware-efficient exact unsigned $4\times4$-bit multiplier for AMD Xilinx 7-series FPGAs. By reorganizing the logic functions mapped to the LUTs, the proposed design uses 11 LUTs and two CARRY4 blocks and achieves a maximum operating frequency of 361.402~MHz. Among the eight evaluated configurations, it has the lowest LUT count and the second-highest maximum operating frequency.
}

\section{Introduction}

With the proliferation of the Internet of Things (IoT) and edge computing, there is growing demand for arithmetic circuits that deliver near-real-time, high throughput under tight budgets on power, silicon area, and memory bandwidth. Unlike large cloud-side accelerators, edge platforms operate under strict power and thermal envelopes, and their latency and responsiveness requirements vary across applications. Consequently, field-programmable gate arrays (FPGAs)—which are programmable yet amenable to specialization that improves energy efficiency—have become a compelling option.

Deep neural network (DNN) inference is a representative FPGA-accelerated workload. Low-bit quantization (e.g., 8-bit or 4-bit) reduces the bit-width of weights and activations, simultaneously lowering memory bandwidth and arithmetic cost and, as a result, enabling both reduced data movement and increased parallelism. On FPGAs, inference engines typically realize such low-precision arithmetic by instantiating many small-granularity multipliers in parallel. The look-up table (LUT) footprint and critical-path delay of each multiplier determine the overall degree of parallelism and the achievable maximum operating frequency ($F_{\max}$). Therefore, even marginal per-multiplier improvements in LUT count or delay can translate into non-negligible system-level gains in performance, area, and power when deploying dense arrays of multipliers.

Although 4-bit and 8-bit multiplications can be implemented using digital signal processing (DSP) blocks via SIMD-style packing, DSP resources impose strong constraints due to their limited count, fixed placement, and the long interconnect they often require. In designs that pack arithmetic units as densely as DNN accelerators, these constraints can lead to routing congestion and $F_{\max}$ saturation. Moreover, heavy use for multiplication can leave insufficient resources for accumulators and for quantization/activation logic. As a result, there remains strong demand for LUT-based low-bit multipliers, and area-efficient, high-performance multipliers are of particular importance.

This work proposes an exact unsigned $4\times4$-bit multiplier with a low LUT count and a high maximum operating frequency. Compared with the evaluated LM implementation of Yao et al.~\cite{Yao2022}, the proposed design reduces the LUT count from 12 to 11 and increases the maximum operating frequency from 296.912 to 361.402~MHz, corresponding to an 8.3\% reduction in LUT count and a 21.7\% increase in maximum operating frequency. The design uses two CARRY4 blocks, compared with one for LM. We evaluate the design on AMD Xilinx 7-series FPGAs.

The remainder of this paper is organized as follows. Section~\ref{sec:preliminary} reviews the relevant FPGA logic architecture. Section~\ref{sec:related} summarizes prior exact multipliers. Section~\ref{sec:proposed} details the structure and operation of the proposed 4-bit multiplier. Section~\ref{sec:results} evaluates hardware resource usage and maximum operating frequency. Section~\ref{sec:conclusion} concludes the paper.

\section{Preliminaries}
\label{sec:preliminary}

\begin{figure}[tb!]
  \centering
  \begin{minipage}{0.9\columnwidth}
    \centering
    \includegraphics[width=0.45\linewidth,pagebox=cropbox]{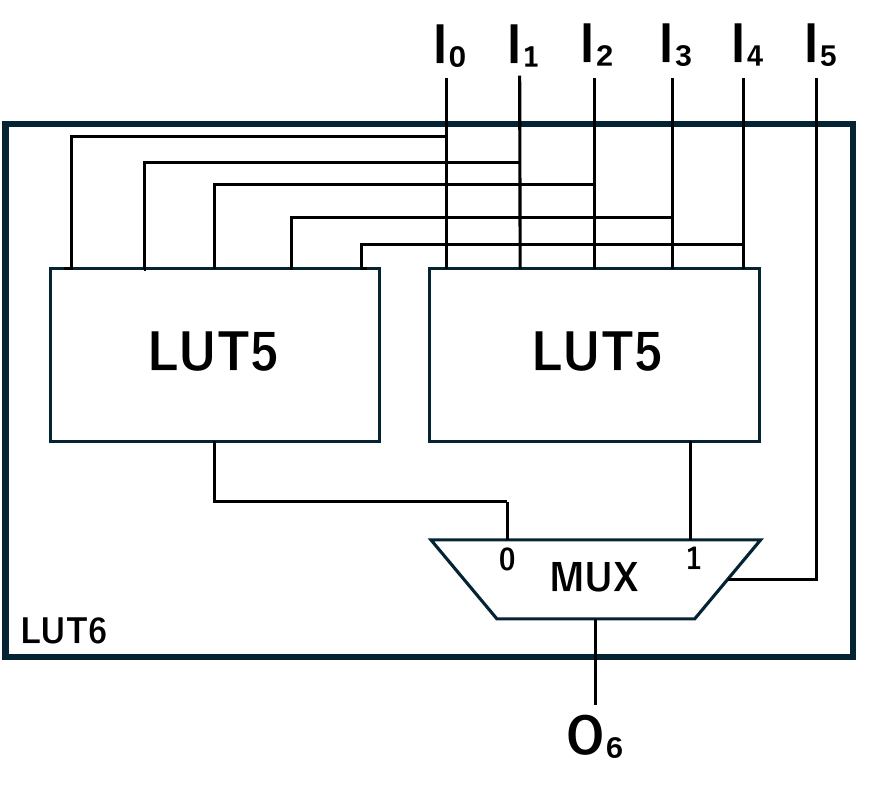}
    \includegraphics[width=0.45\linewidth,pagebox=cropbox]{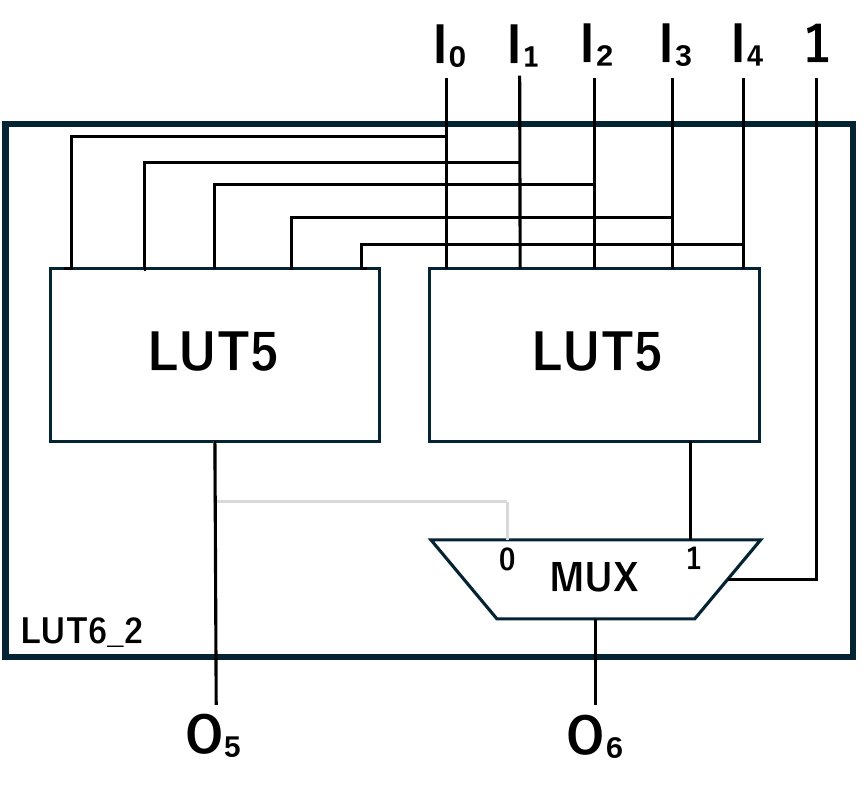}
    \caption{Configuration of LUT6 and LUT6\_2. Xilinx LUTs can be configured to operate as either a 6-input 1-output or a 5-input 2-output device by setting the MUX behavior.}
    \label{fig:LUT6}
  \end{minipage}
\end{figure}

In Xilinx FPGAs, the fundamental logic element is the slice. Each slice contains four 6-input lookup tables (LUTs). Figure~\ref{fig:LUT6} shows the organization of a LUT6 and a LUT6\_2. A LUT ordinarily operates as a 6-input, single-output function (referred to as a LUT6). By controlling the final-stage multiplexer, the same resource can also be used as two 5-input, single-output LUTs that share their inputs; this mode is exposed as the LUT6\_2.

Both LUT6 and LUT6\_2 are provided as synthesizable primitives in Vivado and can be instantiated directly in HDL. In these primitives, the output for every input combination is specified by initializing the truth table (the INIT parameter).

\begin{figure}[t!]
  \centering
  \includegraphics[width=7cm,pagebox=cropbox]{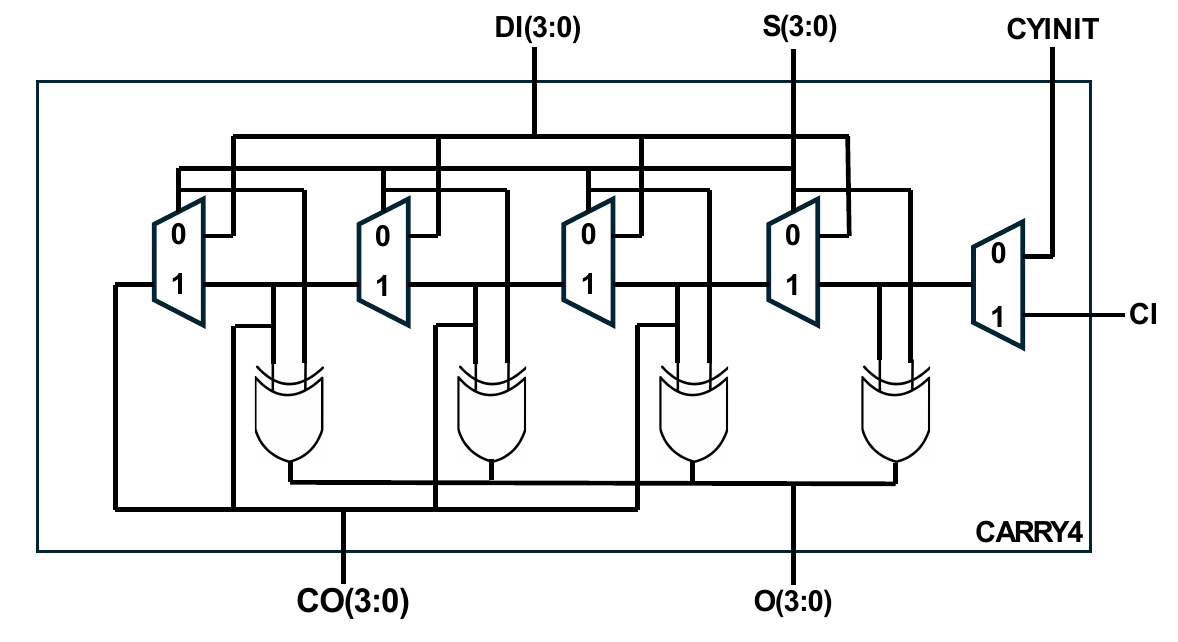}
  \caption{CARRY4 configuration. At each stage, O is the XOR of S and the incoming carry. CO equals the incoming carry when S is 1 and DI when S is 0.}
  \label{fig:CARRY4}
\end{figure}

Each slice includes a CARRY4 block for dedicated carry propagation.
LUT outputs can directly feed this block.
Figure~\ref{fig:CARRY4} shows its structure.
Each CARRY4 block computes four carry bits, allowing addition with
significantly less delay than a carry chain implemented using LUTs alone.
CARRY4 is also available as a synthesizable Vivado primitive
and can be instantiated directly in HDL.

The most significant carry output of a CARRY4, CO[3], can connect to the CI input of the next CARRY4 through dedicated routing. This connection allows CARRY4 blocks to form longer carry chains without routing the inter-block carry through the general interconnect.

\section{Related Work}
\label{sec:related}

This section reviews prior work on 4-bit multipliers for FPGAs.

Yao et al. \cite{Yao2022} proposed 4-bit and 8-bit approximate multipliers for FPGAs and, as the basis of their designs, developed an exact 4-bit multiplier. Their exact design is implemented with 12 LUTs and one CARRY4.

Ullah et al. \cite{Ullah2022} proposed exact and approximate multiplier architectures that exploit FPGA LUTs and dedicated carry chains. The implementation of their exact unsigned 4×4 multiplier evaluated in this work uses 12 LUTs and three CARRY4 blocks, as reported in Table~\ref{tab:hardware}.

Ullah et al. \cite{Ullah2018} proposed a design-space exploration method to generate approximate multipliers of arbitrary bit widths, and—for comparison—also provided a methodology for exact multipliers of arbitrary widths. The 4-bit exact multiplier shown in their paper uses 12 LUTs and three CARRY4s.

Rehman et al. \cite{Rehman2016} proposed an ASIC-oriented architectural exploration framework to generate multipliers with a wide range of approximation levels. In that context, the 4-bit exact multiplier used for comparison is composed of four exact 2-bit multipliers.

Wang et al. \cite{Wang2023} introduced unsigned approximate multipliers that combine the Booth algorithm with probabilistic error correction, achieving substantial area reductions for larger bit-widths. The 4-bit exact multiplier employed in their evaluation uses 13 LUTs and four CARRY4s.

Guo et al. \cite{Guo2024} proposed hardware-efficient FPGA-based approximate multipliers featuring LUT sharing and carry switching, together with a library of 8-bit approximate multipliers supporting multiple multiplication modes. To illustrate the methodology of LUT sharing and carry switching, they show a 4-bit exact multiplier implemented with 13 LUTs and one CARRY4.

All of the above works primarily focus on proposing approximate multipliers. In each case, an exact multiplier is also provided either as a baseline for the approximate designs or as a comparison point. For the LM, Ullah~\cite{Ullah2022}, and LOAM implementations evaluated in Section~\ref{sec:results}, the proposed multiplier achieves a lower LUT count and a higher maximum operating frequency. CARRY4 usage is reported separately in Table~\ref{tab:hardware}.

\section{Proposed Design}
\label{sec:proposed}

\begin{figure}[t]
  \centering
  \includegraphics[width=6cm,pagebox=cropbox]{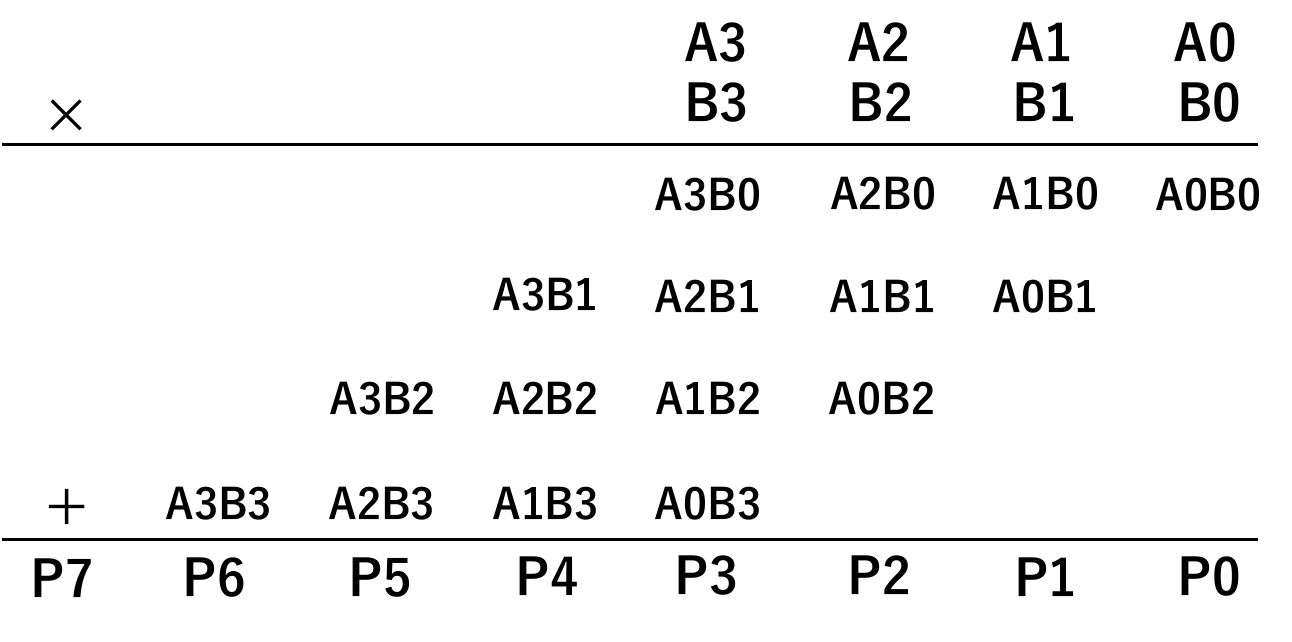}
  \caption{Partial product accumulation for unsigned $4\times4$-bit multiplication.}
  \label{fig:pp}
\end{figure}

\begin{figure}[!t]
  \centering
  \includegraphics[width=\columnwidth,pagebox=cropbox]{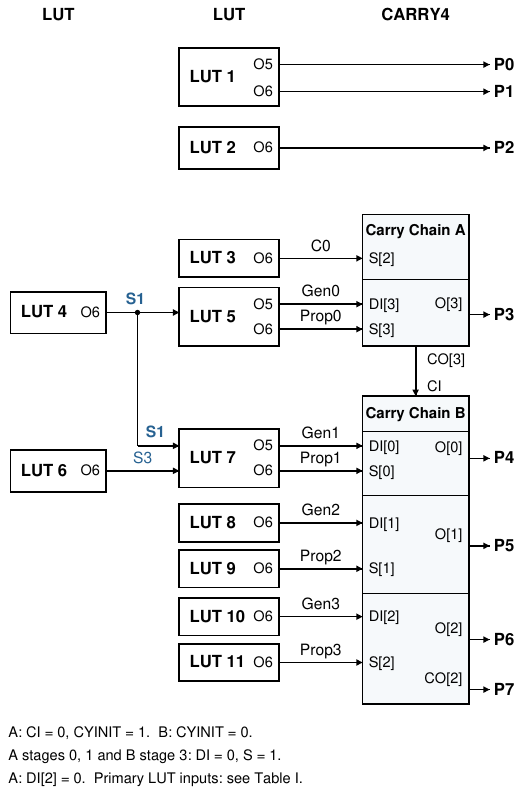}
  \caption{Block diagram of the proposed exact 4-bit multiplier.}
  \label{fig:block_diagram}
\end{figure}

The proposed architecture combines 11 LUTs with two CARRY4 blocks to achieve a low LUT count and a high maximum operating frequency.

Using Fig.~\ref{fig:pp}, we illustrate the unsigned $4\times4$-bit multiplication performed by the proposed design. The multiplicand $A=(A_3,A_2,A_1,A_0)$ and the multiplier $B=(B_3,B_2,B_1,B_0)$ are unsigned integers satisfying $0\leq A,B\leq15$. We generate sixteen partial products (PPs) by ANDing every bit of A with every bit of B. These PPs are then accumulated by addition. The additions proceed column-wise from the least significant bit (LSB) upward while propagating carries. The final product is an unsigned 8-bit word $P=A \times B$.

Fig.~\ref{fig:block_diagram} shows the overall organization of the proposed multiplier. The design uses 11 LUTs in total together with 2 CARRY4 blocks. In the figure, LUTs are numbered 1 through 11, and the CARRY4 blocks are labeled {\it Carry Chain A} and {\it Carry Chain B}.
LUT 4 generates $S_1$, which feeds LUTs 5 and 7, while LUT 6 generates $S_3$, which feeds LUT 7. These connections show the two levels of LUT logic followed by the CARRY4 blocks from left to right. The CARRY4 blocks are arranged vertically, with the product outputs on the right. Primary LUT input connections are omitted for clarity; their assignments are specified in Table~\ref{tab:lutfunction}.

Xilinx FPGAs provide two types of 6-input LUT primitives: a single-output LUT (LUT6) and a dual-output LUT (LUT6\_2). Our implementation uses three dual-output LUTs (LUTs 1, 5, and 7) and eight single-output LUTs. The product bits $P_3$ through $P_7$ are generated using the two CARRY4 blocks.

Table~\ref{tab:lutfunction} summarizes the implementation of the 11 LUTs shown in Fig.~\ref{fig:block_diagram}. The Function column lists the Boolean expressions, including intermediate signals used to define the outputs. The Inputs column specifies the signals in the order $I_0$ through $I_5$, with constant inputs indicated by 1. The Outputs column identifies the signals assigned to ports $O_5$ and $O_6$. LUTs 1, 5, and 7 use both output ports, whereas the remaining LUTs use only $O_6$. The INIT column gives the corresponding hexadecimal configuration values.

\begin{table*}[t]
\centering
\caption{LUT Functions, Input/Output Signals, and INIT Values of the Proposed Exact 4-bit Multiplier}
\label{tab:lutfunction}
\scriptsize
\setlength{\tabcolsep}{2pt}
\renewcommand{\arraystretch}{1.18}
\begin{tabular}{@{}c p{0.49\linewidth} l l l@{}}
\toprule
LUT & Function & Inputs $(I_0,\ldots,I_5)$ & Outputs & INIT \\
\midrule
1 &
$\begin{aligned}[t]
P_0 &= A_0B_0,\\
P_1 &= A_1B_0 \oplus A_0B_1
\end{aligned}$ &
$(A_0,B_1,B_0,A_1,1,1)$ &
$\begin{aligned}[t]O_5&=P_0,\\O_6&=P_1\end{aligned}$ &
\texttt{0x78887888A0A0A0A0} \\
\midrule
2 &
$P_2=A_2B_0\oplus A_1B_1\oplus A_0B_2\oplus(A_0B_1)(A_1B_0)$ &
$(B_2,A_2,B_0,A_0,B_1,A_1)$ &
$O_6=P_2$ &
\texttt{0x653F6AC06AC06AC0} \\
\midrule
3 &
$\begin{aligned}[t]
C_0={}&(A_1B_1)(A_0B_2)\lor(A_2B_0)(A_1B_1)\\
&\lor(A_2B_0)(A_0B_2)\lor(A_0B_1)(A_1B_0)
\end{aligned}$ &
$(A_2,B_0,A_0,B_1,A_1,B_2)$ &
$O_6=C_0$ &
\texttt{0xF8808080C8000000} \\
\midrule
4 &
$S_1=A_1B_2\oplus A_2B_1\oplus(A_1B_1)(A_0B_2)(A_2B_0)$ &
$(A_1,B_2,A_2,A_0,B_1,B_0)$ &
$O_6=S_1$ &
\texttt{0xF878888878788888} \\
\midrule
5 &
$\begin{aligned}[t]
\mathrm{Prop}_0&=(S_1\oplus A_3B_0)\oplus A_0B_3,\\
\mathrm{Gen}_0&=(S_1\oplus A_3B_0)(A_0B_3)
\end{aligned}$ &
$(B_3,A_0,S_1,A_3,B_0,1)$ &
$\begin{aligned}[t]O_5&=\mathrm{Gen}_0,\\O_6&=\mathrm{Prop}_0\end{aligned}$ &
\texttt{0x8778787808808080} \\
\midrule
6 &
$\begin{aligned}[t]
S_2&=A_3B_1\oplus A_2B_2\oplus A_1B_3,\\
C_1&=(A_1B_2)(A_2B_1),\\
S_3&=S_2\oplus C_1
\end{aligned}$ &
$(B_3,A_1,B_1,A_3,B_2,A_2)$ &
$O_6=S_3$ &
\texttt{0x47B7788878887888} \\
\midrule
7 &
$\begin{aligned}[t]
\mathrm{Prop}_1&=S_3\oplus S_1(A_3B_0),\\
\mathrm{Gen}_1&=S_3S_1(A_3B_0)
\end{aligned}$ &
$(B_0,S_1,A_3,S_3,1,1)$ &
$\begin{aligned}[t]O_5&=\mathrm{Gen}_1,\\O_6&=\mathrm{Prop}_1\end{aligned}$ &
\texttt{0x7F807F8080008000} \\
\midrule
8 &
$\begin{aligned}[t]
C_2&=(A_3B_1)(A_2B_2)\lor(A_1B_3)(A_2B_2)\lor(A_3B_1)(A_1B_3),\\
C_3&=S_2C_1,\\
S_4&=A_3B_2\oplus A_2B_3\oplus C_2,\\
\mathrm{Gen}_2&=S_4C_3
\end{aligned}$ &
$(A_2,B_1,B_3,A_1,B_2,A_3)$ &
$O_6=\mathrm{Gen}_2$ &
\texttt{0x8000000000000000} \\
\midrule
9 &
$\mathrm{Prop}_2=S_4\oplus C_3$ &
$(A_2,B_1,B_3,A_1,B_2,A_3)$ &
$O_6=\mathrm{Prop}_2$ &
\texttt{0x37D760A008A0A0A0} \\
\midrule
10 &
$\begin{aligned}[t]
C_4&=(A_3B_2)(A_2B_3)\lor(A_3B_2)C_2\lor(A_2B_3)C_2,\\
\mathrm{Gen}_3&=(A_3B_3)C_4
\end{aligned}$ &
$(B_2,B_1,A_3,A_1,A_2,B_3)$ &
$O_6=\mathrm{Gen}_3$ &
\texttt{0xE0A0800000000000} \\
\midrule
11 &
$\mathrm{Prop}_3=A_3B_3\oplus C_4$ &
$(A_2,B_1,B_2,A_1,B_3,A_3)$ &
$O_6=\mathrm{Prop}_3$ &
\texttt{0x175F8080A0000000} \\
\bottomrule
\end{tabular}
\end{table*}

In the expression for LUT 6, $C_1$ is an intermediate carry term used to compute $S_3=S_2\oplus C_1$. Before simplification, the carry expression is
\[
\begin{aligned}
C_1={}&(A_1B_2)(A_2B_1)\\
&\lor(A_1B_2)(A_1B_1)(A_0B_2)(A_2B_0)\\
&\lor(A_2B_1)(A_1B_1)(A_0B_2)(A_2B_0).
\end{aligned}
\]
Each of the last two terms implies the first term, so the expression simplifies to $C_1=(A_1B_2)(A_2B_1)$. This simplification allows the carry term to be incorporated into the logic that produces $S_3$.

The final four-stage carry chain can be mapped to a single CARRY4, with $P_3$ through $P_6$ obtained from O[0] through O[3] and $P_7$ from CO[3]. However, when $P_7$ is connected to an output flip-flop (FF), Vivado 2024.2 automatically inserts an additional CARRY4 in our implementation flow. To avoid this automatic insertion, the proposed design explicitly distributes the four arithmetic stages across two CARRY4 blocks. As shown in Fig.~\ref{fig:block_diagram}, the first arithmetic stage is assigned to stage 3 of Carry Chain A, and the remaining three to stages 0--2 of Carry Chain B. The final product bit $P_7$ is taken from CO[2] of Carry Chain B, avoiding the use of CO[3] as the final product output.

In Fig.~\ref{fig:block_diagram}, the labels Gen and Prop denote signals connected to the DI and S inputs of the corresponding CARRY4, respectively. The signals $\mathrm{Gen}_0$ and $\mathrm{Prop}_0$ in Table~\ref{tab:lutfunction} are connected to DI[3] and S[3] of Carry Chain A, whereas $\mathrm{Gen}_1$--$\mathrm{Gen}_3$ and $\mathrm{Prop}_1$--$\mathrm{Prop}_3$ are connected to DI[0]--DI[2] and S[0]--S[2] of Carry Chain B, respectively. The signal $C_0$ is connected to S[2] of Carry Chain A. The CO[3] output of Carry Chain A is connected to the CI input of Carry Chain B.

Carry Chain A is initialized with CI = 0 and CYINIT = 1. Its stages 0 and 1 have DI = 0 and S = 1, while stage 2 has DI = 0 and S = $C_0$. These settings produce CO[2] = $C_0$, supplying the carry input to the first arithmetic stage at stage 3. Carry Chain B has CYINIT = 0 and receives its carry input from CO[3] of Carry Chain A.

In Vivado 2024.2, synthesizing an RTL multiplication expression (e.g., $p=a\times b$) does not produce the proposed structure. The manually designed multiplier uses fewer LUTs than both automatically synthesized Exact implementations. It achieves a higher maximum operating frequency than Exact (Area), while Exact (Perf) achieves a higher maximum operating frequency at the cost of a larger LUT count.

\section{Experimental Results}
\label{sec:results}

We compare the proposed multiplier with implementations from prior work, a multiplier synthesized directly from an RTL multiplication expression, and a Vivado multiplier IP core. We evaluate these designs in terms of resource utilization and maximum operating frequency.

All multipliers under evaluation are described in Verilog HDL and synthesized and implemented using Vivado 2024.2. The designs target the Artix-7 35T FPGA on the Arty A7 board. Functional correctness is verified by exhaustive simulation over all input combinations, confirming that each design computes the product exactly.

For each design, $F_{\max}$ is calculated from static timing analysis after placement and routing as
\begin{equation}
  F_{\max}~[\mathrm{MHz}] = \frac{1000}{T_{\mathrm{target}}-\mathrm{WNS}},
  \label{eq:fmax}
\end{equation}
where $T_{\mathrm{target}}$ is the target clock period used for implementation and WNS is the worst negative slack reported for that clock, both expressed in nanoseconds.

The evaluation includes the proposed multiplier, denoted as “Proposed,” and three types of baselines. \mbox{“Exact”} denotes a multiplier synthesized directly from an RTL multiplication expression. \mbox{“Vivado IP”} denotes a multiplier generated using the multiplier IP in Vivado 2024.2. We also evaluate selected exact multipliers from the studies reviewed in Section~\ref{sec:related}.

{\raggedright
The Perf configuration uses the Flow\_\allowbreak PerfOptimized\_\allowbreak high synthesis strategy and the Performance\_\allowbreak ExploreWithRemap implementation strategy. The Area configuration uses the Flow\_\allowbreak AreaOptimized\_\allowbreak high synthesis strategy and the Area\_\allowbreak Explore implementation strategy. Proposed and the evaluated implementations from prior work (LM, Ullah, and LOAM) use Perf. Exact and Vivado IP are each evaluated with both Perf and Area.\par
}

\begin{table}[!htbp]
\caption{Resource Utilization and Maximum Operating Frequency of 4-bit Multipliers}
\label{tab:hardware}
\centering
\footnotesize
\setlength{\tabcolsep}{4pt}
\renewcommand{\arraystretch}{1.15}
\begin{tabular}{@{}lccr@{}}
\toprule
Multiplier & LUTs & CARRY4 & $F_{\max}$ (MHz) \\
\midrule
\textbf{Proposed} & 11 & 2 & 361.402 \\
LM~\cite{Yao2022} & 12 & 1 & 296.912 \\
\cite{Ullah2022} & 12 & 3 & 311.139 \\
LOAM~\cite{Guo2024} & 13 & 1 & 336.587 \\
Exact (Area) & 15 & 2 & 326.052 \\
Exact (Perf) & 20 & 2 & 382.995 \\
Vivado IP (Area) & 15 & 4 & 290.276 \\
Vivado IP (Perf) & 20 & 4 & 296.824 \\
\bottomrule
\end{tabular}
\end{table}

\begin{samepage}
Table~\ref{tab:hardware} compares resource utilization and maximum operating frequency ($F_{\max}$). Resource utilization is reported as the number of LUTs and CARRY4 primitives, while performance is reported as $F_{\max}$ in MHz. The comparison includes the proposed design, LM~\cite{Yao2022}, the exact multiplier of Ullah et al.~\cite{Ullah2022}, LOAM~\cite{Guo2024}, Exact, and Vivado IP.
\par
\end{samepage}

The proposed multiplier uses 11 LUTs and two CARRY4 blocks and achieves an $F_{\max}$ of 361.402~MHz. It has the lowest LUT count and the second-highest $F_{\max}$ among the eight evaluated configurations. Compared with Exact (Perf), which achieves 382.995~MHz using 20 LUTs, it uses 45.0\% fewer LUTs with a 5.6\% lower $F_{\max}$, while using the same number of CARRY4 blocks.

\begin{figure}[t]
  \centering
  \includegraphics[width=8.5cm,pagebox=cropbox]{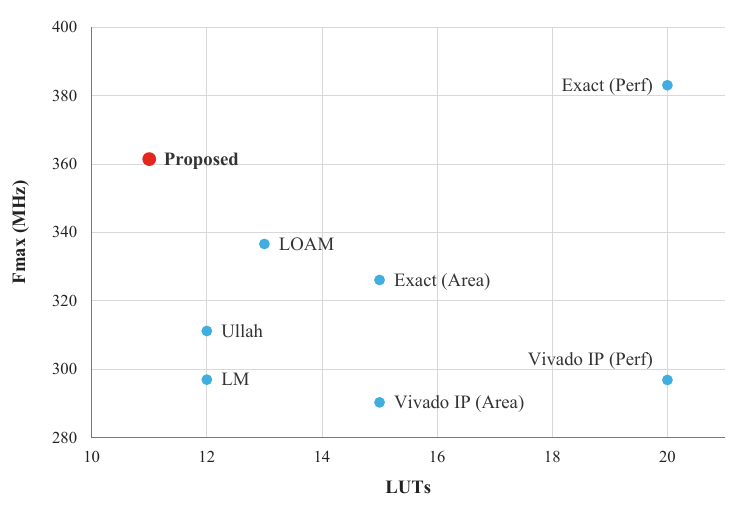}
  \caption{LUT count and maximum operating frequency of 4-bit multipliers.}
  \label{fig:performance}
\end{figure}

Fig.~\ref{fig:performance} illustrates the trade-off between LUT count and $F_{\max}$ for the configurations in Table~\ref{tab:hardware}. Designs toward the upper left combine fewer LUTs with a higher $F_{\max}$. Within the evaluated set, the proposed design and Exact (Perf) are Pareto-optimal with respect to these two metrics: no other evaluated configuration improves one metric without worsening the other. The proposed design provides the lowest LUT count, whereas Exact (Perf) provides the highest $F_{\max}$.

\section{Conclusion}
\label{sec:conclusion}

This paper presented an exact unsigned $4\times4$-bit multiplier for AMD Xilinx 7-series FPGAs. By reorganizing the logic functions mapped to the LUTs, the proposed design uses 11 LUTs and two CARRY4 blocks and achieves a maximum operating frequency of 361.402~MHz. Among the eight evaluated configurations, it achieves the lowest LUT count and the second-highest maximum operating frequency. Compared with Exact (Perf), it reduces the LUT count by 45.0\% with a 5.6\% reduction in maximum operating frequency.

\begingroup
\footnotesize
\setlength{\emergencystretch}{1em}

\bibliographystyle{unsrt}
\bibliography{library}
\par
\endgroup

\end{document}